\documentclass[11pt,twoside]{article}

\usepackage{asp2014}

\aspSuppressVolSlug
\resetcounters

\begin{document}

\title{A crude but efficient pipeline for JWST MIRI imager : the case of sn1987A}
%\title{Home Made Pipeline for MIRI Imager}

% Note the position of the comma between the author name and the 
% affiliation number.
% Authors surnames should come after first names or initials, eg John Smith, or J. Smith.
% Author names should be separated by commas.
% The final author should be preceded by "and".
% Affiliations should not be repeated across multiple \affil commands. If several
% authors share an affiliation this should be in a single \affil which can then
% be referenced for several author names.
% See ManuscriptInstructions.pdf and ASPmanual2010.pdf 3.1.4 for more details

\author{R.~Gastaud,$^1$ and A.~Coulais$^{2,1}$ 
\affil{$^1$Universit\'e Paris-Saclay, Universit\'e Paris Cit\'e, CEA, CNRS, AIM, Gif-sur-Yvette, France \email{Rene.Gastaud@cea.fr}\\
  $^2$LERMA, Observatoire de Paris, Universit\'e PSL, Sorbonne Universit\'e, CNRS, Paris, France \email{Alain.Coulais@obspm.fr}}}

% This section is for ADS Processing.  There must be one line per author.
\paperauthor{R. Gastaud}{Rene.Gastaud@cea.fr}{0009-0007-5200-1362}
            {CEA}{IRFU DAp}{Saclay Gif-sur-Yvette}{}{91191}{France}

\paperauthor{A. Coulais}{Alain.Coulais@obspm.fr}{0000-0001-6492-7719}
            {Observatoire de Paris}{LERMA}{Paris}{}{75014}{France}

% There should be one \aindex line (commented out) for each author. These are used to
% build up the author index for the Proceedings. The surname must come first, followed by
% initials. Note the use of ~ before each initial to control spacing.
% The \author entries (see above) have surname last. These \aindex entries have
% surname first.

% keep the %% !
%\aindex{Gastaud,~R.}
%\aindex{Coulais,~A.}

\begin{abstract}

Most of the space projects or large observatories do have official
tools like simulators, end-to-end pipelines developed during years by
a large team of contributors. They are like {\em cathedrals}.  In this
paper, we show that very simplistic code using basic operators provided
by high level language like GDL allows to write quickly high quality
code to process raw data into scientifically validated outputs. This is
{\em bazaar}.

In this paper we argument why we consider large infrastructure should
be designed to allow small ones to benefit from it and allow to graft
better alternative processing at very low cost.
  
\end{abstract}

\section{introduction}

Observations in imager mode of supernova sn1987A with MIRI have been
made in July 2022, just after Performance Verification Phase of the
instruments onboard space mission Webb (aka "James Webb Space
Telescope" or JWST) \citep{Wright2023}. We quickly realized that the
official pipeline was too limited at that time and does not give the
high quality level we need.  So we developed in IDL/GDL a simple but
efficient pipeline to process our data from scratch.  Being able to
code with our favorite language (GDL : 
\citet{Park2022,CoulaisADASS2024}) was a way to quickly test new ideas
and have high quality results, without waiting for official product.

\section{What can we expect from a pipeline ?}

What is a pipeline ?  A pipeline ingests the data from any instrument,
any observation mode, and processes it by steps to a meaningfully
science data.  Pipeline automatically runs on all data, do not confuse
with Data Analysis Tools which requires science decisions (human
interaction), but a good pipeline is flexible and can be used as a Data
Analysis Tool.

What is a good pipeline : this depends upon the user. An average user
wishes to have:
\begin{itemize}
  \setlength\itemsep{-0.25em}
  \item  easy and quick installation on a personal computer
  \item good documentation, with a lot of examples (notebook or python
    scripts)
  \item speed and memory : could be run on a personal computer on
    reasonable amount of time
  \item input/output for each stage/step should be stored in
    self-documented open-format files containing information on the
    previous steps, with the name of the calibration files used, and
    the version of the pipeline.
\end{itemize}

An advanced user wishes more:
\begin{itemize}
  \setlength\itemsep{-0.25em}
  \item modularity: each stage/step can be individually called
  \item flexibility: each step has default method and parameters which
    can be over-ridden
  \item  parametrable log facility
  \item  documentation on each step (methods and parameters)
\end{itemize}

A programmer wishes even more: 
\begin{itemize}
  \setlength\itemsep{-0.25em}
  \item unit test for each stage/step
  \item each step can be done by different methods, easily substituable
  \item use of a configuration management for the code (eg Git or SVN)
  \item documentation on the unit tests, heritage of each module,
    architecture, etc...
  \item use of  external libraries (do not reinvent the wheel)
\end{itemize}

\begin{figure}[!ht]
  \includegraphics[width=0.45\linewidth]{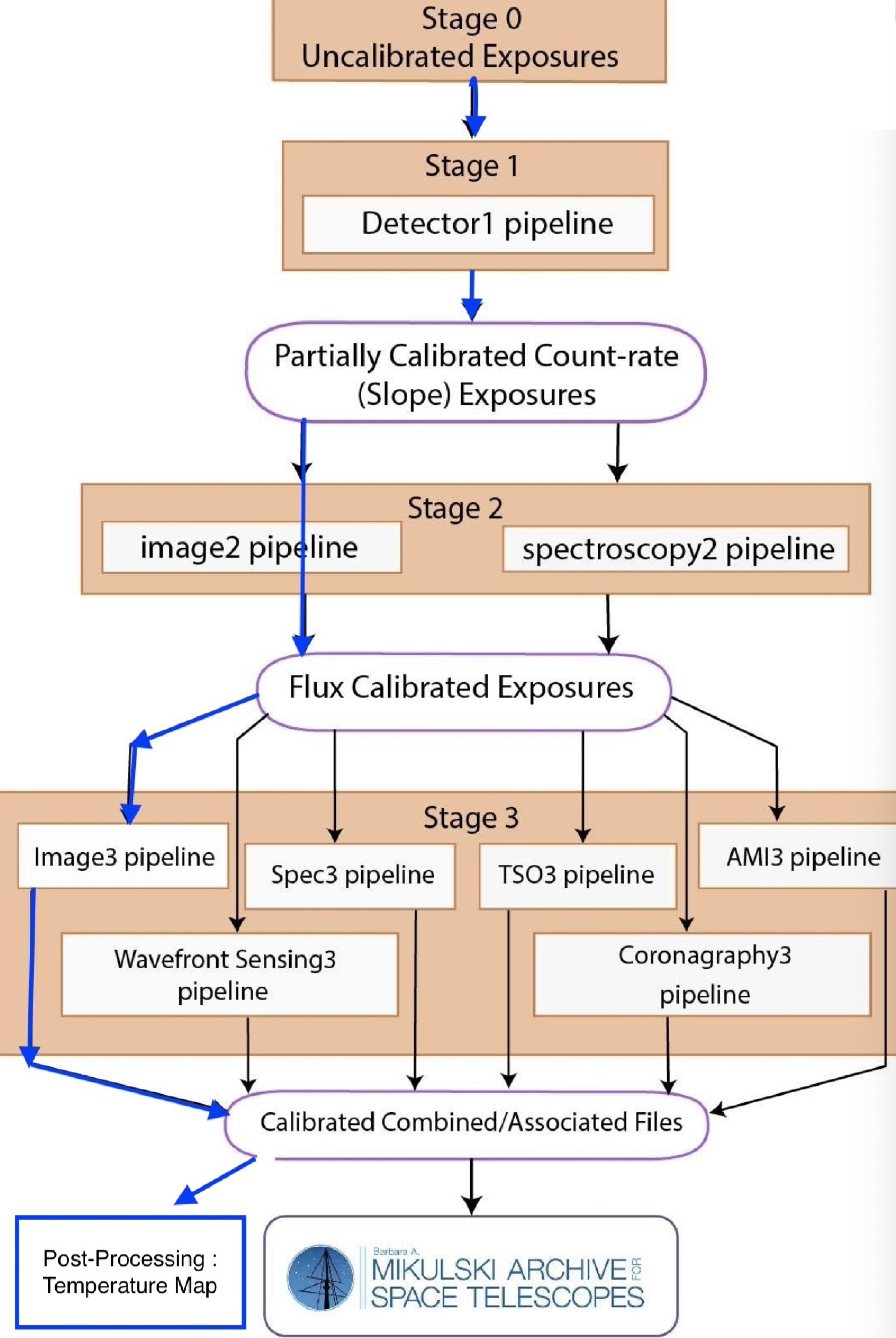}
  \includegraphics[width=0.45\linewidth]{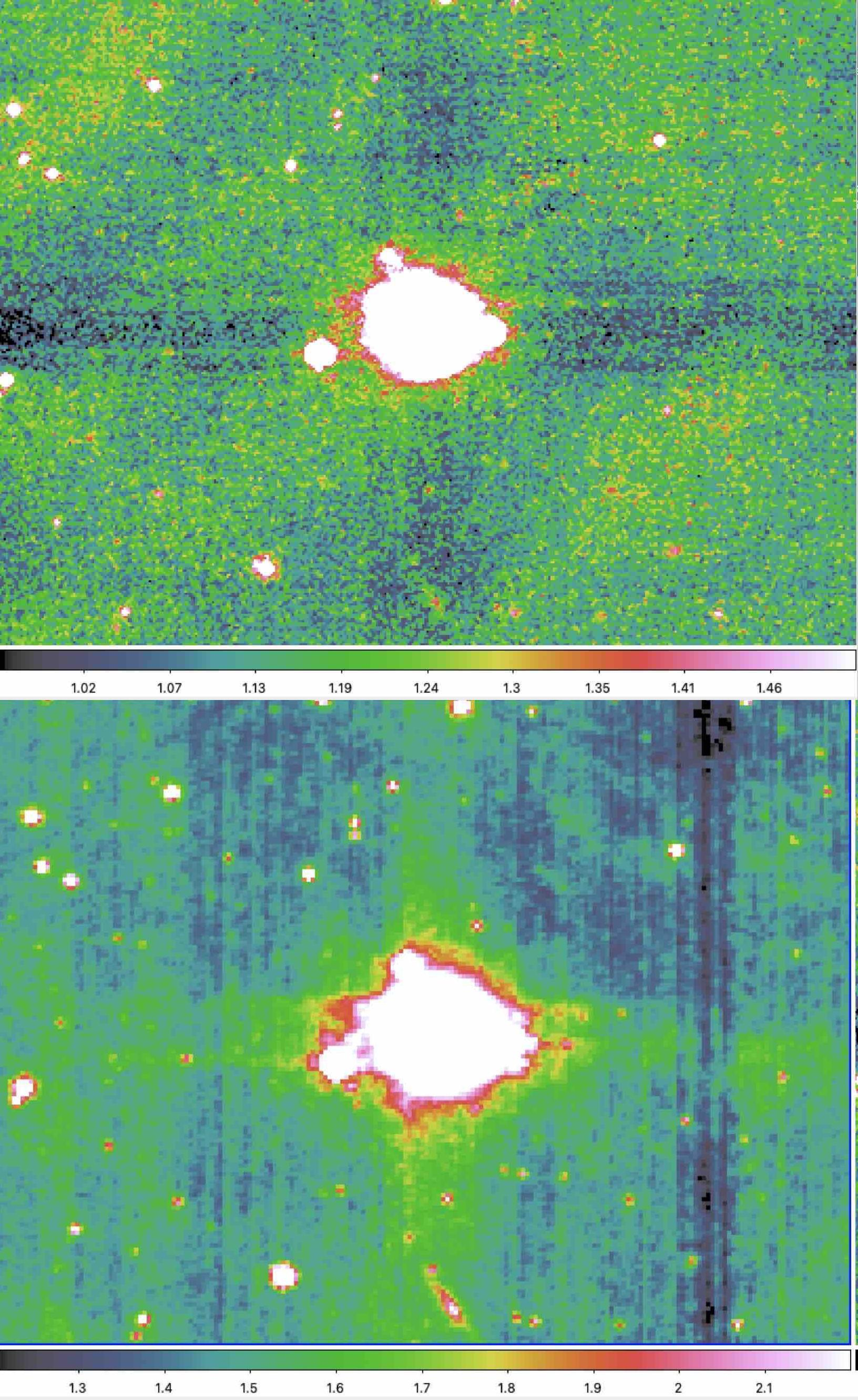}
  \caption{Left : JWST pipeline, with in blue the home-made. Right : mosaicking
    images : top home-made, bottom official pipeline}
\end{figure}

\section{JWST Official Pipeline}

\citet{Gordon2015} described the official JWST MIRI pipeline
It respects most of the points we list in the previous section.  We can
discuss that the documentation is very limited, and the examples
too. We were able to locate only a very limited set of unitary tests,
poorly documented, and we don't find end-to-end tests.

JWST is an observatory coming with four instruments, each one are a
combination of a large number of observing modes and optical
configurations. For MIRI, we have LRS, MRS, imager, full or sub-array
modes, etc...  The installation of the official pipeline comes with
all the modes, takes time, requires Conda and a lot of disk
space. Even if you have very limited data (eg a small LRS observation)
all the code for all the 4 instruments and all the modes are installed
and calibrations products are downloaded. At the end, before running
the code, the total amount of space is surprisingly large (tens of Go).

FITS files haves a long story in the ADASS conferences.  In the JWST
case, the input/output of each step, intermediate or final, and the
calibration data are stored in FITS files.  These files contain the
detailed information of the previous steps, the version of the
pipeline and also the version of the associated calibration files
(CDP).  The very interesting point to use FITS files is it can be
read back by most of the tools used in astronomy (ds9, fv, IDL/GDL, ...).

The pipeline is divided in 3 main stages. Each stage produces an output
which data has different dimensions and or different unis, so this
division is not arbitrary. Each stage is divided in several steps,
which can be turned on/off, whose methods and parameters can be
selected.  A stage or a step can be executed alone. The input/output of
each step can be stored (not the default option).  This is great to
tune the pipeline to your data, the problem is again the documentation
(some name can be misleading).

The pipeline is versatile : the same code for the different
instruments as much as possible, both for storing data and processing
data. This is possible with inheritance. The drawback is a code more
difficult to understand.  The pipeline uses as much as possible
existing routines/libraries, for example the astropy package.

%%%%%%%%%%%%%%%%%%%%%%%%%%%%%%%%%%%%%%%%
\section{Dedicated Home-made Pipeline}

At the beginning of JWST scientific life, we had to process an
observation of SN1987 made by the MIRI Imager \citep{Bouchet2024}.
This observation was done with four different filters, and with a
subarray 512x512 to avoid saturation (the sampling time is quicker
than the full array).  Four different exposures with different
pointing were taken for each filter (Dithering) to reduce the impact
of bad pixels.  We simplify the steps and avoid to use the
official calibration products : some were not available, some were not
optimal (in our case), and some were not up to date.  Our home-made
pipeline starts from raw data and can be used only one mode of the
observatory (see blue line in figure 1 left).

We made significant improvements in data processing for
\begin{itemize}
  \setlength\itemsep{-0.25em}
  \item ramps to slopes, rewriting a method with a better Allan
    Variance (we use median filtering)
  \item odd/even removal (less SNR on individual maps) : we remove
    dark subtraction step which only adds noise and compute a better
    background
  \item mosaicking (better resolution in maps) we rebin of a factor 2
    to take into account the half-pixel shifts
    \end{itemize}
When compared to the initial official product, clearly the
revisited one was better, which was {\em a must} for the
post-processing done in \citet{Bouchet2024}.

Even if those improvements are now known and published, some are still
not in the official pipeline. Thanks to our experience as data
scientists, following KISS philosophy, we wrote the custom tools
working on raw data, circumventing the constraints from official
infrastructure, in a month.

As usual, the end of the pipeline is not the end of the processing.
We have an extra-step which we call post-processing (see figure 1).
In our case, from the 4 maps each at differnts wavelenght (5.6, 8, 12,
15 um) we want to compute a temperature map. This was easily written
in GDL using the astrolib (blackbody and MPFIT \citep{Markwardt2009}).

\section{Conclusion}

In the ADASS context we discuss what can be a good pipeline nowadays,
enriched from the experience of JWST and other spatial projects.

JWST official pipeline was an excellent tool because without having to
learn internal details of the official pipeline, we can extract data
from it, or insert data into it, which allows very quick comparisons
between alternatives methods for given steps, with alternative
language.  Then in few weeks and few simple routines we propose three
methods (ramp processing, odd/even removal and mosaicking) better than
the official ones.

\bibliography{p914}

\acknowledgements

AC warmly thanks POL and CEA for travel funding. AC thanks RG for
daily usage of GDL since years !

\end{document}